\documentclass[10.5pt]{iopart}

\pdfoutput=1
\usepackage{geometry}                % See geometry.pdf to learn the layout options. 
\usepackage{iopams,amsfonts,amssymb,amsthm} 
\usepackage[colorlinks=true,citecolor=blue,linkcolor=red,urlcolor=blue]{hyperref}
\usepackage[usenames,dvipsnames,svgnames]{xcolor}
\usepackage[font=footnotesize, labelfont=it, margin=0.5cm]{caption}
\usepackage{rawfonts}
\usepackage{graphicx}

\usepackage{threeparttable}
\usepackage{tablefootnote}
\usepackage{multirow}

\def\sfrac#1#2{\scalebox{1}{$\frac{#1}{#2}$}}

\def\lfl{\lfloor} \def\rfl{\rfloor} \def\lcl{\lceil} \def\rcl{\rceil}

\def\Ref#1{(\ref{#1})}

\input prepictex
\input pictex
\input postpictex

\usepackage[font=footnotesize,labelfont=sf]{caption}
\begin{document}

\title[A star polymer escaping from a pore]{\textsf{The escape transition 
of a lattice star polymer grafted in a pore}}

\author{CJ Bradly$^1\dagger$ \& EJ Janse van Rensburg$^2\ddagger$}

\address{$^1$School of Mathematics \& Statistics,
The University of Melbourne, Victoria 3010, Australia}
\address{$^2$Department of Mathematics \& Statistics, 
York University, Toronto, Ontario M3J~1P3, Canada}
\ead{$\dagger$\href{mailto:chris.bradly@unimelb.edu.au}{chris.bradly@unimelb.edu.au}
$\ddagger$\href{mailto:rensburg@yorku.ca}{rensburg@yorku.ca}}
\vspace{10pt}
\begin{indented}
\item[]\today
\end{indented}

\begin{abstract}
Polymers in confined spaces are compressed and have reduced conformational
entropy, and will partially or fully escape from confinement if conditions are suitable.  
This is in particular the case for a polymer grafted in a pore.  The escape of the polymer
from the pore may be considered a partial translocation from the pore into bulk solution, 
and the resulting conformational readjustment of the polymer has characteristics 
of a thermodynamic phase transition.  In this paper a lattice self-avoiding walk model 
of a star polymer grafted in a pore is examined numerically using the PERM algorithm.  
We show that the arms of the grafted lattice star escape one at a time as the length 
of the pore is reduced, consistent with earlier results in the literature.  Critical points 
for the escape transitions are estimated for square and cubic lattice models and we 
also examine various properties of the model as it undergoes the escape transition.
\end{abstract}

%
% Uncomment for keywords
\vspace{0pc}
\noindent{\it Keywords}:  Escape transition, star polymer \\

%\pacs{82.35.Lr,82.35.Gh,61.25.Hq}
\ams{82B41,82D60,82M31}

%\maketitle

\section{Introduction}

Grafted and adsorbed polymers have important applications in the modern
world, including the targeted and timed delivery of drugs using polymer 
coatings on medical devices, such as stents \cite{CVZ10,JM12,SYS15}, or
in nano-particle and polymer systems \cite{SWP96}.  Other applications 
include the stabilization of colloid dispersions by adsorbing polymers 
on colloid particles \cite{LTR78,F85,WP86,NDA97}, the adsorption 
of biopolymers at membranes \cite{H18}, or their translocation 
through pores \cite{SK03}.  These are all instances of systems where the 
properties of polymers, grafted on a hard wall, or in a confining space,
or adsorbed in confining spaces, are important.  Here the changes in
conformational entropy of the polymers when confined is an important
determinator of its properties, a fact known since the pioneering work
of PJ Flory \cite{F49,F56,F69} and other polymer scientists more than 70 years 
ago \cite{E65,F66,deG79}.

In addition to the above, atomic force microscopy (AFM) made the manipulation 
of single polymer molecules possible \cite{H99,ZZ03}.  A polymer grafted to 
a substrate and underneath the tip of the AFM is compressed, and it may 
escape from confinement by rearranging its conformations to expand into
bulk \cite{GWS97}.  In this ensemble the polymer is said to be compressed
by a piston (the AFM tip) onto an anvil (the substrate to which the polymer is
grafted at an endpoint) and then to \textit{escape} from the confining space.
This escape is a phase transition, corresponding to a critical point in the phase
diagram for the polymer.  The properties and scaling of the escape transition 
were examined numerically in references  \cite{GWS97,MYB99, HBKS07,PMEB13}.  
Escape transitions of the branches in compressed or confined star polymers 
were simulated in references \cite{S00,RC15}.  Amongst these models are 
lattice models \cite{HBKS07,JvR23}, bead-spring models \cite{MYB99}, as 
well as molecular dynamics calculations \cite{PMEB13}.  

Phenomenological approaches to the escape transition of a polymer
confined underneath a piston based on a ``blob analysis'' \cite{deG79} 
models the free energy as a function of the separation between the
piston and anvil confining the polymer \cite{GWS97}.  The escape transition 
is seen at a critical separation and appears to be a first order transition 
\cite{MYB99,HBKS07} (consistent with the numerical results in reference 
\cite{JvR23}), but it was also argued to have unconventional properties, for
example negative compressibility \cite{DKSKB09}.  The transition was also 
modelled as a function of a force conjugate to the separation between 
the piston and anvil, and numerical data show that the transition occurs 
at a critical force \cite{MYB99}, although the order of the transition in this
case may be uncertain \cite{HBKS07}.

\begin{figure}
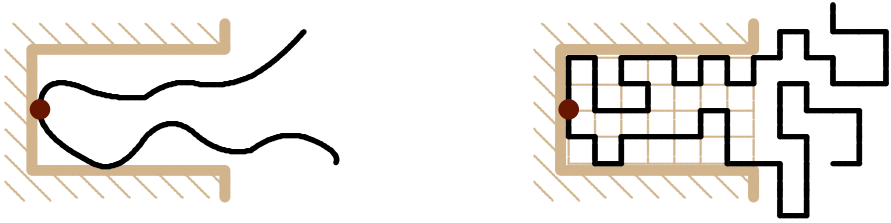

    \beginpicture
\setcoordinatesystem units <1pt,1pt>
\setplotarea x from -70 to 0, y from -10 to 80
\startrotation by 0 -1 about 20 35

\color{Tan}
\setplotsymbol ({\scalebox{1.0}{$\bullet$}})
\plot -13 70 -3 70 -3 -3 43 -3 43 70 53 70 /
\setplotsymbol ({\scalebox{0.2}{$\bullet$}})
\plot -13 60 -3 70 /  \plot -13 50 -3 60 /  \plot -13 40 -3 50 / \plot -13 30 -3 40 /
\plot -13 20 -3 30 /  \plot -13 10 -3 20 / \plot -13  0 -3 10 /  \plot -13 -10 -3 0 /
\plot -3  -13 7 -3 /  \plot  7 -13 17 -3 /  \plot  17 -13 27 -3 /  \plot  27 -13 37 -3 /
\plot  37 -13 54 4 /  \plot  47 -13 54 -6 /  \plot  43 4 53 14 /  \plot  43 14 53 24 /
\plot  43 24 53 34 /  \plot  43 34 53 44 /  \plot  43 44 53 54 /  \plot  43 54 53 64 /
\plot  43 64 48 69 /

\color{black}
\setplotsymbol ({\scalebox{0.5}{$\bullet$}})
\setquadratic
\plot 20 0 30 5 40 20 40 30 30 40 25 50 30 60 35 70 35 80 30 90 30 100
35 110 40 112 /
\plot 20 0 10 5 13 20 15 30 15 40 10 50 10 60 10 70 7 80 0 90 -10 100  /
\setlinear

\color{Sepia}
\put {\scalebox{2.0}{$\bullet$}} at 20 0
\stoprotation

\setcoordinatesystem units <1pt,1pt> point at -200 0 
\setplotarea x from -10 to 70, y from -10 to 80
\startrotation by 0 -1 about 20 35
\color{Tan}
\setplotsymbol ({\scalebox{0.2}{$\bullet$}})
\plot 0 0 0 70 40 70 40 0 0 0 /  \plot 10 0 10 70 /  \plot 20 0 20 70 /
\plot 30 0 30 70 /  \plot 0 10 40 10 /  \plot 0 20 40 20 /  \plot 0 30 40 30 /
\plot 0 40 40 40 /  \plot 0 50 40 50 / \plot 0 60 40 60 /

\setplotarea x from -10 to 70, y from 40 to 80
\setplotsymbol ({\scalebox{1.0}{$\bullet$}})
\plot -13 70 -3 70 -3 -3 43 -3 43 70 53 70 /
\setplotsymbol ({\scalebox{0.2}{$\bullet$}})
\plot -13 60 -3 70 /  \plot -13 50 -3 60 /  \plot -13 40 -3 50 /  \plot -13 30 -3 40 /
\plot -13 20 -3 30 /  \plot -13 10 -3 20 /  \plot -13  0 -3 10 /  \plot -13 -10 -3 0 /
\plot -3  -13 7 -3 /  \plot  7 -13 17 -3 /  \plot  17 -13 27 -3 /  \plot  27 -13 37 -3 /
\plot  37 -13 54 4 /  \plot  47 -13 54 -6 /  \plot  43 4 53 14 /  \plot  43 14 53 24 /
\plot  43 24 53 34 /  \plot  43 34 53 44 /  \plot  43 44 53 54 /  \plot  43 54 53 64 /
\plot  43 64 48 69 /

\color{black}
\setplotsymbol ({\scalebox{0.5}{$\bullet$}})
\plot 20 0 30 0 30 10 40 10 40 20 30 20 30 30 30 40 30 50 20 50
20 60 30 60 40 60 40 70 40 80 50 80 60 80 60 90 50 90 40 90 
30 90 30 80 20 80 10 80 10 90 20 90 20 100 20 110 30 110 40 110
40 100 /
\plot 20 0 10 0 0 0 0 10 10 10 20 10 20 20 20 30 10 30 10 20 
0 20 0 30 0 40 10 40 10 50 0 50 0 60 10 60 10 70 0 70 0 80 -10 80
-10 90 0 90 0 100 10 100 10 110 10 120 0 120 -10 120 -10 110
-10 100 -20 100 /

\color{Sepia}
\put {\scalebox{2.0}{$\bullet$}} at 20 0
\stoprotation

\color{black}\normalcolor
\endpicture
\caption{(Left) A schematic diagram of a $2$-star grafted in a pore.
The two arms of the star explore conformation inside and outside
the pore.  If the pore is deep, then the star is completely retracted 
inside the pore.  If the pore is shallow, or the $2$-star has long arms, 
then one or both arms escape from the pore and explore conformations 
in bulk outside the pore. (Right) A square lattice model of a $2$-star 
grafted in a pore.  The star is grafted in the bottom of the pore at
its central node, and its arms are self-avoiding walks exploring 
conformations inside the pore, and in the bulk lattice to the
right of the pore in this diagram.}
\label{F1}
\end{figure}

Related to the escape transition of polymers confined underneath a piston, 
there are also a substantial literature devoted to the study of polymers
escaping when confined in nanoscale pores \cite{SLV00,DMB06}.   In these 
studies polymers grafted in pores and their scaling properties were 
examined.   If the pore is short with an open end, or if the polymer is long, 
then there is the possibility that the polymer escapes from the pore at a 
critical length \cite{M05}, a phenomenon closely related to the translocation 
of polymers through nano-channels \cite{SK03} and with polymer escape from
underneath a piston.  In reference \cite{M05} the escape of a grafted polymer 
from a pore is examined.  Using Flory-Huggins style arguments to model 
the free energy, a first order escape transition was proposed, consistent with 
later numerical work on lattice and other models in references 
\cite{DKSKB09,JvR23}.  In this paper we generalise the models to include
grafted star polymers escaping from a pore, as illustrated schematically
in figure \ref{F1}.    

\begin{figure}
\includegraphics[width=0.475\textwidth,height=0.3\textheight]{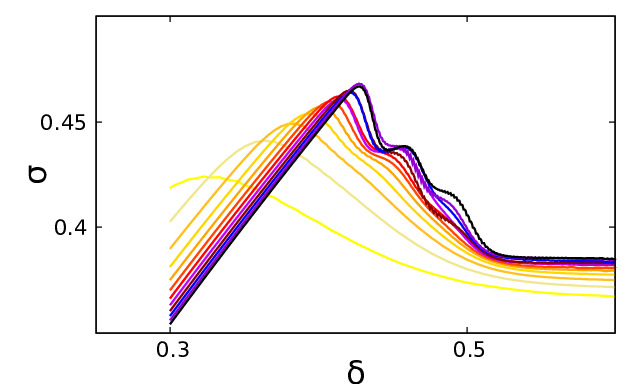}
\includegraphics[width=0.475\textwidth,height=0.3\textheight]{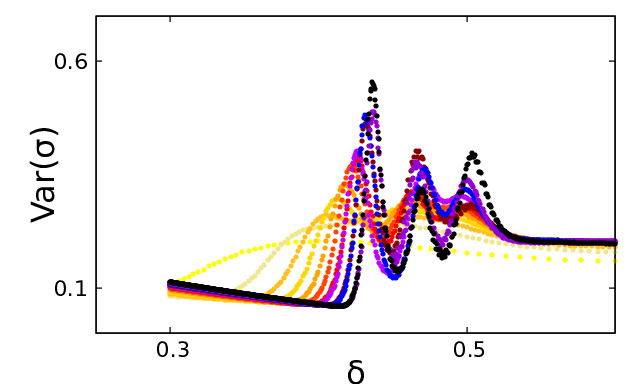}
\caption{(Left)  The total span $\sigma$ (sum of the spans of the individual arms)
of cubic lattice $3$-stars escaping from a $3\times 3\times L$ pore.  The hue of 
the colours increases from yellow to black as $L$ increases in steps of $10$
from $10$ to $120$.  Undulations in the curves on the left show changes in 
the total span as the arms escape from the pore with changes in the relative 
depth $\delta$ of the pore.  
(Right) The variance of the total span plotted as a function of the relative depth
$\delta$ of the pore.  With increasing relative depth $\delta$ (the ratio of the 
length of the arms of the star to the (fixed) depth of the pore) three distinct
peaks develop in the variance, consistent with three separate escape transitions
(one for each arm).  The most prominent peak (also the left-most peak) corresponds
to the primary escape transition, while the other peaks to its right correspond to
secondary escape transitions.}
\label{F1a}
\end{figure}

There are numerical evidence that the arms of a star polymer escape one
at a time when the polymer is compressed by a piston \cite{PMEB13}.  In this
situation there would be an escape transition for each arm of the star, at different
critical points, since the escape of one arm vacates space underneath the piston
and so has an impact on the conformational entropy of the other arms.
In this paper we investigate this phenomenon for a simpler model, namely a 
model of a star polymer with its central node grafted in the bottom of a cylindrical
pore and with arms that can (partially) escape from the pore.  Our numerical 
results will show that the arms in our model do indeed escape one-by-one, 
as the length of the pore is reduced.  Moreover, the primary escape transition 
has a strong first order character (see, for example, references \cite{M05,PMEB13}), 
but the subsequent escape transitions (even if they are first order as well) have 
somewhat weaker signatures in the metric and thermodynamic
quantities of the model.  For example, the total span (namely the sum of the spans
of the individual arms along the direction of the pore) of a $3$-star grafted in the
pore is plotted as a function of relative pore depth $\delta$ in the left panel
in figure \ref{F1a}.  If the pore is shallow (that is, when $\delta$ is small)
then all the arms have escaped, and they explore conformations (partially) 
in bulk, with spans along the pore increasing as the pore becomes deeper 
(with increasing $\delta$), since the arms are stretched inside the pore.  
At critical values of the relative pore depth $\delta$ there are undulations 
in the total span curves.  These correspond to the escapes of each of the 
three arms and the variance of the total span peaks sharply at each of the
undulations as shown in the right panel of figure \ref{F1a}.  The most prominent
of the peaks in the variances is the left-most (peaking at the smallest value of
$\delta$ compared to the other peaks).  This peak corresponds to the last arm to
escape as the pore depth decreases (or the first arm to retract into
the pore as the pore depth increases).  Since this peak is consistently the most
prominent, we consider it to be the \textit{primary or dominant escape transition}, 
while the subsequent peaks correspond to \textit{secondary escape transitions}
of the remaining arms.

The \textit{primary arm} of the star corresponds to the primary escape transition
and it is also the last escape transition to occur when the relative depth of the pore
is decreased.  The primary arm escapes when all the all the other arms of the 
star are already exploring conformations outside the pore.  If an arm is not 
the primary arm, then it is a denoted a \textit{secondary arm} and it is
associated with one of the secondary escape transitions.

The outline of the paper is as follows.  In section \ref{sec:LatticeStarsModel} 
we define the model and discuss briefly the PERM algorithm \cite{G97} 
and its implementation \cite{PK04,CJvR20} to collect data for self-avoiding 
walk stars of total length $10L$ in a pore of depth $L\in\{10,20,30,\ldots,120\}$.

In section \ref{sec:NumericalResults} we analyse our data for $2$-stars escaping
from a pore in the square lattice.  Consistent with figure \ref{F1a}, our data
for this case also show a strong primary escape transition, consistent with a
first order transition, and a weaker secondary escape transition.  We determine
the location of the critical points (as shown in table \ref{t1}) by examining the
location of peaks in the second derivative of the free energy (this is the
\textit{pressure gradient} in the model), and also the peaks in the total span of the
star, the fraction of arms inside the pore, as well as the variance of this last
quantity.  This gave four estimates of the locations of the critical points, 
separated by a few statistical confidence intervals.  We combined the estimates
by taking a simple average, and then taking half the difference between 
the largest and the smallest estimates as a confidence interval.  This gives
$\delta_c = 0.698(11)$.  This error bar is slightly larger than the statistical error 
bars estimated, and we take it as our confidence interval on the estimate of the 
primary escape transition of square lattice $2$-stars shown in table \ref{t1}.
The location of the secondary transition was similarly estimated.

By examining the heights of the peaks in the pressure gradient, a finite size 
crossover exponent $\phi_m=0.97(12)$ is estimated for the primary escape 
transition, consistent with this transition being first order.  The large error bar 
in this estimate is due to strong corrections to scaling and uncertainties in our data.  

Cubic lattice data for $2$- and $3$-stars were also collected (see 
section \ref{sec:CubicStars}).  In these cases we were able to extract good estimates
of the location of the critical points using the pressure gradient, but data collected
on the total span (see figure \ref{F1a}) or fraction of the arms inside the pore 
were more noisy.  The best estimates on the location of critical points stated 
in table \ref{t1} were obtained from the pressure gradient data, and the stated error
bars are double the statistical confidence intervals of these estimates. 

\begin{table}[h!]
\caption{Critical points for escape transitions of lattice $f$-stars in a pore}
\begin{indented}
\lineup
\item[]
\begin{tabular}{l@{}*{3}{c}ccccc}
%\hline          
Model & Pore width $w$ & $\delta_c^{(1)}$ & $\delta_c^{(2)}$ & $\delta_c^{(3)}$ \cr
\hline
\vspace{-2mm}
& & & & & \cr
square lattice & \multirow{2}{*}{$3$} & \multirow{2}{*}{$0.698(11)$} 
                                                    & \multirow{2}{*}{$0.756(30)$} \cr
$2$-stars  \cr
\hline
\vspace{-2mm}
& & & & & \cr
\multirow{1}{*}{cubic lattice} \cr
\multirow{1}{*}{$2$-stars} 
          & $3$ & $0.4140(66)$ & $0.4565(62)$ & $ $ & $ $ \cr
\multirow{1}{*}{$3$-stars} 
          & $3$ & $0.448(13)$ & $0.478(11)$ & $0.512(17)$ & $ $ \cr
\hline
%\vspace{-1mm}
%& & & & & \cr
%\multicolumn{3}{l}{$\dagger$ -- Comment} \\
\end{tabular}
\end{indented}
\label{t1}
\end{table}

In addition to the results shown in table \ref{t1}, we also considered the mean
total spans and the fractions of arms retracted into the pore in our models in
sections  \ref{sec:NumericalResults} and \ref{sec:CubicStars}.  The results
are consistent with the data presented for the free energies of the models, and 
provide additional support for the first order nature of the primary escape
transition and (perhaps) also for the secondary transitions.  Our results show
that when the arms are retracted that they are (on average) completely contained 
within the pore.  This gives rise to a rebounding effect in our data -- as one arm 
escapes, it vacates space in the pore, with the result that there 
are more conformational degrees of freedom available for itself, and for the
other (still retracted) arms.  This is seen in an ``overshoot phenomenon'' when the
lengths of the arms within the pore is plotted against the relative pore depth.

In section \ref{sec:WeightedSpan} the effects of a pulling force on the escape 
transition is generally examined by weighing the total span of $3$-star 
conformations in the cubic lattice with a fugacity $y=e^{F/kT}$ where $F$ 
can be interpreted as a pulling force (and $T$ is the absolute temperature and
$k$ is Boltzmann's constant in lattice units).  Plotting phase diagrams by plotting 
the mean total span and its variance against $(\delta,y)$ then show curves of 
escape transitions corresponding to the escape of each arm for a wide range 
of values of the fugacity $y$.  A pulling force on the endpoints of the arms
will turn them ballistic in the direction parallel to the pore direction
(see, for example, references \cite{JvRW16,JvRW18,JvRW22} for more on 
the effects of a pulling force on a grafted self-avoiding walk or lattice star).  
However, this applied force does not cause the immediate escape of the 
arms of the star, which, for fixed relative pore depth $\delta$ remains 
retracted until the force passes critical values where the arms escape, 
again one-at-a-time.  This is not an unexpected observation, and in a 
general sense the effects of external forces, solvent quality, interactions 
between the star polymer and the walls of the pore, temperature, and 
other interactions, and the location and nature of the escape transition, 
remain largely unexplored in the context provided in this paper.

We conclude the paper with a few final remarks in section \ref{Conclusion}.

\section{Lattice stars grafted in a pore}
\label{sec:LatticeStarsModel}

\begin{figure}[h]
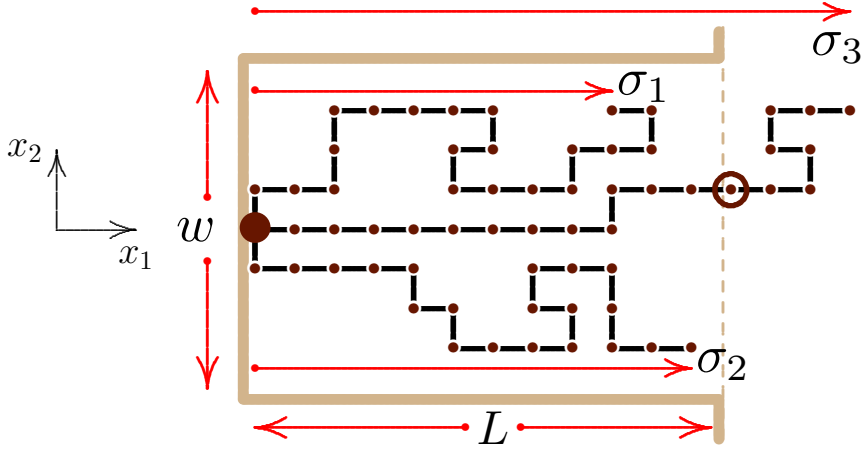

    \beginpicture
    \setcoordinatesystem units <1.5pt,1.5pt>
    \setplotarea x from -70 to 200, y from -10 to 100
    
    \color{Black}
    \arrow <10pt> [.2,.67] from -50 40 to -30 40
    \arrow <10pt> [.2,.67] from -50 40 to -50 60
    \put {\Large$x_1$} at -30 33  \put {\Large$x_2$} at -58 60

    \color{Tan}
    \setplotsymbol ({\scalebox{1.0}{$\bullet$}})
    \plot 117 -13 117 -3 -3 -3 -3 83 117 83 117 90 / 
    \setplotsymbol ({\scalebox{0.25}{$\bullet$}})
    \setdashes <4pt>
    \plot 118 -13 118 93 /
    \setsolid
    
    \color{Black}
    \setplotsymbol ({\scalebox{0.5}{$\bullet$}})
    \plot 0 40 0 50 10 50 20 50 20 60 20 70 30 70 40 70 50 70 60 70 
    60 60 50 60 50 50 60 50 70 50 80 50 80 60 90 60 100 60 100 70 
    90 70 /
    \plot 0 40 10 40 20 40 30 40 40 40 50 40 60 40 70 40 80 40 90 40
    90 50 100 50 110 50 120 50 130 50 140 50 140 60 130 60 130 70
    140 70 150 70 /
    \plot 0 40 0 30 10 30 20 30 30 30 40 30 40 20 50 20 50 10 60 10 
    70 10 80 10 80 20 70 20 70 30 80 30 90 30 90 20 90 10 100 10 
    110 10 /
    
    \color{White}
    \multiput {{\scalebox{1.5}{$\bullet$}}} at
     0 40 0 50 10 50 20 50 20 60 20 70 30 70 40 70 50 70 60 70 
    60 60 50 60 50 50 60 50 70 50 80 50 80 60 90 60 100 60 100 70 
    90 70 /
    \multiput {{\scalebox{1.5}{$\bullet$}}} at
    0 40 10 40 20 40 30 40 40 40 50 40 60 40 70 40 80 40 90 40
    90 50 100 50 110 50 120 50 130 50 140 50 140 60 130 60 130 70
    140 70 150 70 /
    \multiput {{\scalebox{1.5}{$\bullet$}}} at
    0 40 0 30 10 30 20 30 30 30 40 30 40 20 50 20 50 10 60 10 
    70 10 80 10 80 20 70 20 70 30 80 30 90 30 90 20 90 10 100 10 
    110 10 /
    
    \color{Sepia}
    \multiput {{\scalebox{1.0}{$\bullet$}}} at
     0 40 0 50 10 50 20 50 20 60 20 70 30 70 40 70 50 70 60 70 
    60 60 50 60 50 50 60 50 70 50 80 50 80 60 90 60 100 60 100 70 
    90 70 /
    \multiput {{\scalebox{1.0}{$\bullet$}}} at
    0 40 10 40 20 40 30 40 40 40 50 40 60 40 70 40 80 40 90 40
    90 50 100 50 110 50 120 50 130 50 140 50 140 60 130 60 130 70
    140 70 150 70 /
    \multiput {{\scalebox{1.0}{$\bullet$}}} at
    0 40 0 30 10 30 20 30 30 30 40 30 40 20 50 20 50 10 60 10 
    70 10 80 10 80 20 70 20 70 30 80 30 90 30 90 20 90 10 100 10 
    110 10 /
    
    \circulararc 360 degrees from 120 54 center at 120 50 
    
    \color{Sepia}
    \put {\scalebox{3.0}{$\bullet$}} at 0 40 
    
    \color{Red}
    \setplotsymbol ({\scalebox{0.25}{$\bullet$}})
    \arrow <10pt> [.2,.67] from 0 5 to 110 5
    \arrow <10pt> [.2,.67] from 0 75 to 90 75
    \arrow <10pt> [.2,.67] from 0 95 to 150 95
    
    \arrow <12pt> [.2,.67] from 67 -10 to 115 -10
    \arrow <12pt> [.2,.67] from 53 -10 to 0 -10
    
    \arrow <12pt> [.2,.67] from -12 48 to -12 80 
    \arrow <12pt> [.2,.67] from -12 32 to -12 0 
    \multiput {\scalebox{0.75}{$\bullet$}} at 0 5 0 75 0 95 
    -12 48 -12 32 67 -10 53 -10 /
    
    \color{Black}
    \put {\scalebox{2.0}{$\sigma_1$}} at 98 76
    \put {\scalebox{2.0}{$\sigma_3$}} at 147 86
    \put {\scalebox{2.0}{$\sigma_2$}} at 118 6
    
    \put {\scalebox{2.0}{$w$}} at -15 40 
    \put {\scalebox{1.8}{$L$}} at 60 -10

    \color{black}
    \normalcolor
    
    \endpicture
\caption{A $3$-star in a pore.  The pore has depth $L$ and width
$w$.  The star is grafted near the middle of the bottom of the pore, and
its three arms explore conformations inside the pore, and may escape
into the bulk space by exiting the pore to the right.  The total length $n$ of the star
is the sum of the lengths $(n_1,n_2,n_3)$ of its arms, and it is (almost) 
uniform, so that the arms have almost the same lengths:
$\lfloor n/3\rfloor \leq n_i \leq \lceil n/3 \rceil$ for $1\leq i \leq 3$.
The \emph{spans} of the arms are denoted by $(\sigma_1,\sigma_2,
\sigma_3)$, and clearly $0 \leq \sigma_i \leq \lceil n/3 \rceil$.  The
arms are labelled such that $\sigma_1\leq\sigma_2\leq\sigma_3$.  The 
total span of the star is the sum $\sigma=\sigma_1+\sigma_2+\sigma_3$.  
In each arm there may be a first vertex where it escapes the pore.  
In the figure the third arm escapes the pore, and the first vertex 
outside the pore is indicated by $O$. The other arms do not escape, but are 
retracted inside the pore.}
\label{F2}
\end{figure}

In figure \ref{F2} a schematic diagram of a square lattice $3$-star grafted 
in a pore is shown.  Standard Cartesian coordinates $(x_1,x_2)$ are shown in the
figure.  The pore has depth $L$ in the $x_1$-direction, and width $w$
in the $x_2$-direction.  By introducing a third normal coordinate $x_3$, 
the pore generalises to the cubic lattice, where the width $w$ is replaced 
by a cross-sectional base area in the $x_2x_3$-plane. Normally, this will be 
assumed to be a square of side-length $w$.  One may generalise this to
$d$-dimensions by assuming that the pore has a base area of size $w^{d-1}$
normal to the $x_1$ direction, and length or depth $L$ in the $x_1$ direction. 
In this paper we shall only consider models with $d=2$, or $d=3$.

The $3$-star in figure \ref{F2} is shown with its central node grafted in or near 
the midpoint of the bottom of the pore.   More generally, one may consider
an $f$-star instead, with $f\geq 1$. The arms of the star explore conformations 
which may escape from the pore into the bulk outside the pore. The total 
length of the star is denoted by $n$, and the $f$ arms has lengths
$(n_1,n_2,\ldots,n_f)$.  The star is \textit{almost uniform} in that $\sum_i n_i=n$
and $\lfloor n/f \rfloor \leq n_i \leq \lceil n/f\rceil$ for $1\leq i \leq f$.

\subsection{The free energy and spans of $f$-stars in a pore}

The \textit{span} $\sigma_i$ of the $i^{th}$ arm is the distance along 
the pore to the endpoint of the arm, as shown in figure \ref{F2}. 
If $\sigma_i > L$, then the $i$-th arm is said to have \textit{escaped} 
from the pore, otherwise $\sigma_i\leq L$ and the arm is \textit{retracted} 
in the pore.  We follow the convention that arms are labelled such that 
the sequence of spans $[\sigma_i]=(\sigma_1,\sigma_2,\ldots,\sigma_f)$ 
is non-decreasing.  The \textit{total span} of the $f$-star is denoted 
$\sigma=\sum_i \sigma_i$.  For a given non-decreasing sequence of spans 
$[\sigma_i]$ let $S^{(f)}_n([\sigma_i];w,L)$ be the number of grafted $f$-stars
with spans $[\sigma_i]$ in a pore of length $L$ and width $w$, of total 
length $n$.  Summing over sequences $[\sigma_i]$ with fixed total span $\sigma$
gives the number of stars of length $n$ and total span $\sigma$, denoted
$S^{(f)}_n(\sigma;w,L)$.  The total number of these $f$-stars of 
arbitrary total span is given by
\begin{equation}
S^{(f)}_n(w,L) = \sum_{\sigma=0}^n S^{(f)}_n(\sigma;w,L) 
= \sum_{\sigma=0}^n {\sum_{[\sigma_i]}}^{\Large\prime} S^{(f)}_n([\sigma_i];w,L) ,
\label{eqn1}
\end{equation}
where the primed summation is over all $[\sigma_i]$ such that
$\sum\sigma_i = \sigma$.

The limiting free energy of this model is defined by
\begin{equation}
\mathcal{F}_{w,L} = \limsup_{n\to\infty} \frac{1}{n} \log S^{(f)}_n(w,L) .
\label{eqn2}
\end{equation}
If $w=o(n)$ and $L=o(n)$ then the methods of quarantining arms to
wedges, following the methods of references \cite{HTW82}, will show that
this limit exists (and is equal to $\log \mu_d$ where $\mu_d$ is the
growth constant of the self-avoiding walk \cite{SS93}).

The situation is more interesting if $L=\lfl \delta n / f\rfl$ for some
fixed $\delta>0$ (and with $w$ fixed and for $f$-stars).  Here, one fixes
$\delta$, and then vary $n$ and $L$ proportionally to each other.
This defines a pore of depth proportional to the length of the star.
In the limit as $n\to\infty$, this gives the free energy as a function of $w$ 
and the relative depth $\delta$ of the pore by
\begin{equation}
\xi_{w}^{(f)}(\delta) = \limsup_{n\to\infty} 
\frac{1}{n} \log S^{(f)}_n(w,\lfl \delta n /f \rfl) .
\label{eqn3}
\end{equation}
It is not known that this limit exists. The finite size approximation
to this free energy is
\begin{equation}
F_w^{(f)}(\delta) = \frac{1}{n} \log S^{(f)}_n(w,\lfl \delta n /f \rfl) 
\label{eqn4}
\end{equation}
where for given $L$, one puts $\delta_n = Lf/n$ so that 
$\delta_n \leq \delta < \delta_n+1/n$. This defines 
$F_w^{(f)}(\delta)$ as a step-function.  For example, 
if $1\leq n \leq 100$ and $f=2$, with $L=10$, then $\delta_n = 20/n$
as $n$ increases from $1$ to $100$.  With increasing $n$,
$\delta_n$ decreases, and this decreases the relative depth of the pore.

\subsection{Parallel PERM simulations of escaping lattice stars}

Lattice stars were sampled in the pores by growing them using the PERM 
algorithm \cite{G97}.  The model is shown in figure \ref{F2}.  First the pore 
is defined in the $d$-dimensional hypercubic lattice by its width $w$ and depth $L$.
The \textit{bottom of the pore} has $x_1$-coordinate equal to zero.  The 
$f$-stars are grafted in the pore so that the central node of the star is 
situated at the \textit{central bottom node} of the pore which has 
coordinates $(0,\lfl w /2 \rfl,\ldots ,\lfl w /2 \rfl)$. That is, if $w$ is odd 
then it is off-set from the middle of the bottom of the pore.  PERM samples 
stars by growing $f$ arms from this site.  The boundary of the pore is ensured 
by enforcing that all vertices with first coordinate $0 \leq x_1 \leq L$ (within 
the pore) have other coordinates $0 \leq x_j \leq w$ with $2\leq j \leq d$.
Vertices outside the pore ($x_1 > L$) have no restriction on the other 
coordinates.

The implementation was of a flat histogram sampling method \cite{PK04} 
using Parallel PERM \cite{CJvR20}.  $2$-Stars and $3$-stars were sampled in 
pores of width $w=3$ and depths $L\in\{10,20,30,\ldots,120\}$.  Each pore 
defined by $L$ and $w$ (and dimension) requires a separate simulation in 
which stars are sampled up to total length $N=10L$.  The output raw data 
of each simulation are estimates of $S_n^{(f)}(\sigma;w,L)$, with flat histogram 
sampling of total length $n \leq N$ and total span $\sigma \leq N$.  This gives 
estimates of the finite size free energy as shown in equation \Ref{eqn4}.  
The simulations proceeded over $T$ parallel PERM sequences sampling states 
in $B$ blocks (and each block of length $M$ started PERM tours).  This gives a 
total of $TBM$ tours.  In each case $T=4$ (and in a few cases $T=6$),
while we fixed $B=250$ and $M=10{,}000$ tours.  This gives a total of 
$10{,}000{,}000$ PERM tours for each value of $L$.

\begin{figure}
\includegraphics[width=0.5\textwidth,height=0.3\textheight]{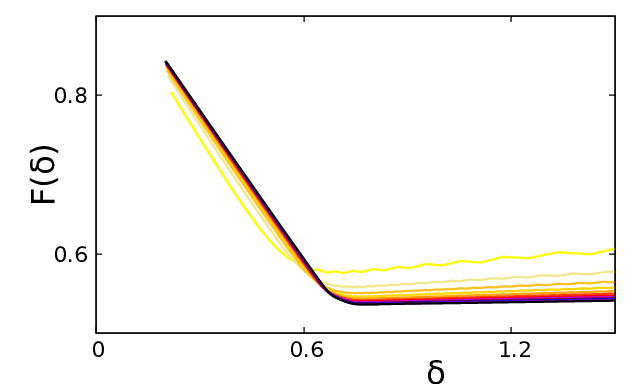}
\includegraphics[width=0.5\textwidth,height=0.3\textheight]{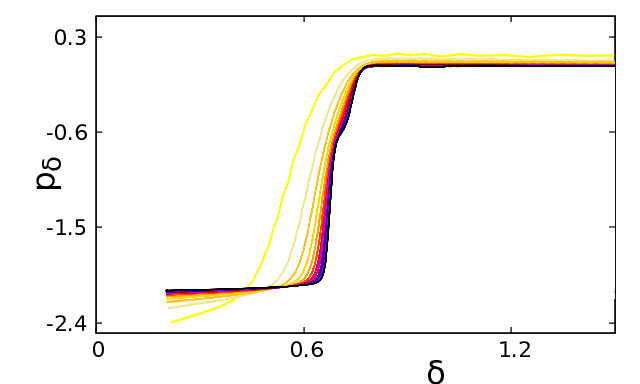}
\caption{(Left panel) The free energy per unit length of square lattice $2$-stars 
in a pore of width $w=3$.  The colour increases from yellow to black as the pore 
depth $L$ increases in the sequence $(10,20,\ldots,120)$.  The maximum 
length of the arms is $10L$, and $\delta=2\,L/n$ is the relative depth of
the pore as a fraction of the length $n/2$ of an arm in the $2$-star.   With 
increasing $L$ the curves converge to a limiting curve with a clear transition 
at a critical value $\delta_c$ of $\delta$.  When $\delta>\delta_c$ (and for 
large $L$) the free energy approaches a constant independent of $\delta$.
When $\delta<\delta_c$ the free energy increases quickly with decreasing
$\delta$.  (Right panel) The derivative $p_\delta=\sfrac{d}{d\delta} F_{w}^{(f)}$
(the entropic pressure) in the pore as a function of $\delta$.  For large $\delta>\delta_c$ 
the arms of the $2$-star are retracted inside the pore, and the pressure is
positive and approaches zero with increasing $L$.  
When $\delta<\delta_c$ the pressure becomes large negative, 
showing that the arms of the star escape from the pore.  Notice the minor 
deviation of the curves close to the critical point indicating a more complex
transition.}
\label{F3}
\end{figure}

\section{Numerical results}
\label{sec:NumericalResults}

In this section we consider the case of 2-stars grafted in a square lattice pore 
with width $w=3$ in order to establish the numerical methods used.
In later sections we will apply these methods to some variations of the model.

\subsection{Free energy}

The finite size free energies $F_w^{(f)}(\delta)$ (see equation \Ref{eqn4})
are plotted against the relative pore depth $\delta = Lf/n$ with $f=2$
in figure \ref{F3} (left panel) for $2$-stars in a square lattice pore of width 
$w=3$.   The total length $n$ of the star takes values $0\leq n \leq 10 L$ as
the pore depth $L$ increases in $\{10,20,30,40,\ldots,120\}$ in increments of
$10$.  The colours of the curves increase in hue from yellow via red to black with 
increasing $L$.  As $L$ increases, the curves appear to approach a limiting free 
energy (given by equation \Ref{eqn3}).  Escape transitions of the star, or parts 
of the stars, correspond to non-analyticities in the limiting free energy 
$\xi_w^{(f)}(\delta)$ at critical values of $\delta$ (the relative pore depth).  
This may be explored by taking (numerical) derivatives of $F_w^{(f)}(\delta)$ 
and plotting those as a function of $\delta$.  There are parity effects in our 
data due to the underlying lattice and maybe due to the fact that our stars
are nearly uniform, as explained earlier. We accommodate for these parity
effects by analysing even and odd labelled points separately.

The first derivative $\sfrac{d}{d\delta} F_w^{(f)}(\delta) = p_\delta$ 
is the change in the free energy to a variation of the pore depth.  This 
corresponds to a (linear) \textit{entropic pressure} $p_\delta$ in the pore, 
and it should be non-decreasing with increasing relative pore depth $\delta$ 
since the star retracts into the pore as its depth increases.  This is shown in 
the right panel of figure \ref{F3}, where the curves increase quickly near 
a critical value of $\delta$.  The curves appear to approach a limiting shape 
which is constant for $\delta$ small, or for $\delta$ large, and away from 
the critical point(s).  The escape transition is exposed by the quickly increasing 
limiting curve near the critical point(s).  Notice the small variation visible 
in the vertical portion of the curve; this indicates a $2$-step escape transition 
in the model.  

\begin{figure}
\centering
\includegraphics[width=0.67\textwidth,height=0.3\textheight]{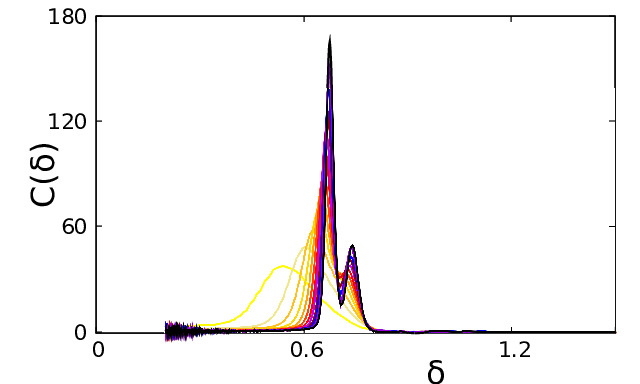}
\caption{The pressure gradient curves as a function of $\delta$, the relative
pore depth.  For small $L$ the curves have a single broadened peak, but the
peak narrows are $L$ increases in $(10,20,\ldots,120)$, and a second
peak, lower in height, appears.  These peaks show that there are conformational
rearrangements of the walks at two critical values of $\delta$, each rearrangement
corresponding to the escape of an arm from the pore.  With decreasing
$\delta$ the first arm escapes at the lower peak, and then the second arm,
and so the rest of the star, escapes at the higher peak.  The prominent
(higher) peak corresponds to the primary transition, while the lower peak
peak is a secondary transition.}
\label{F4}
\end{figure}

The second derivative $\sfrac{d^2}{d\delta^2}  F_w^{(f)}(\delta)
= \sfrac{d}{d\delta} p_\delta$ can be considered to be related to a linear 
(and isothermal) compressibility $\kappa_\delta$ by defining 
\[ C(\delta) = \sfrac{d^2}{d\delta^2}  F_w^{(f)}(\delta)
= \sfrac{d}{d\delta} p_\delta = 1/\kappa_\delta .\]
The derivative $C(\delta)$ is the inverse of $\kappa_\delta$, and we call it
a \textit{pressure gradient}.  $C(\delta)$ is small and positive if 
$\kappa_\delta$ is large (and so the pore and the star it contains is compressible), 
and large when $\kappa_\delta$ is small (and so when the star is stiff and the pore 
is not compressible).  In figure \ref{F4} the pressure gradient is plotted for 
square lattice $2$-stars when $w=3$.  As before,  the curves are increasing 
in hue as $L$ increases in steps of $10$ from $L=10$ (yellow) to $L=120$ (black).  
As $L$ increases, the pressure gradient develops two peaks (one a prominent 
primary peak, and then a secondary peak off-set to larger values of $\delta$).  
For small values of $L$ the peaks are broadened, and the dominant peak 
absorbs the secondary peak.  The two peaks are well separated when $L\geq 70$.  

In table \ref{t2} the location and height of the peaks in the pressure
gradient curves in figure \ref{F4} are shown.   The peaks correspond to
the escape of arms from the pore, with the primary escape transition corresponding
to the dominant peak on the left, and the secondary escape transition to the 
secondary peak on the right. Thus, in this model a decreasing the relative 
pore depth $\delta$ from a large value first takes the model through an escape 
transition at the secondary peak when the secondary arm escapes, and then 
through a primary escape transition at the dominant peak when the primary 
arm escapes.  We label the peaks from the left so that the $m$-th peak at pore 
depth $L$ by is located at $\delta_L^{(m)}$, and with its height denoted 
by $h_L^{(m)}$.  For example, the primary escape transition is signalled in
the pressure gradient curves by the left-most peak located at $\delta_L^{(1)}$ 
with peak height $h_L^{(1)}$.   

\begin{table}[h!]
\caption{Escape transitions of $2$-stars in a pore of width $3$ and depth $L$}
\begin{indented}
\lineup
\item[]
\begin{tabular}{@{}*{3}{l}{l}llll}
%\hline          
Model & $L$ & $\delta_L^{(1)}$ & $\delta_L^{(2)}$ & $h_L^{(1)}$ & $h_L^{(2)}$  \cr
\hline
\vspace{-2mm}
& & & & & \cr
\multirow{2}{*}{$2$-stars} 
          & $10$ & $0.541(30)$ & $-$ & $37.47(91)$ & $-$ \cr
\multirow{2}{*}{$d=2$} 
          & $20$ & $0.606(19)$ & $-$ & $48.3(1.1)$ & $-$ \cr
\multirow{2}{*}{$w=3$} 
          & $30$ & $0.632(14)$ & $-$ & $58.08(75)$ & $-$ \cr
          & $40$ & $0.640(11)$ & $-$ & $68.5(1.3)$ & $-$ \cr
          & $50$ & $0.6494(85)$ & $-$ & $79.2(1.5)$ & $-$ \cr
          & $60$ & $0.6593(73)$ & $0.7143(86)$ & $90.6(1.9)$ & $33.22(56)$ \cr
          & $70$ & $0.6635(63)$ & $0.7292(76)$ & $102.2(1.8)$ & $35.30(57)$ \cr
          & $80$ & $0.6639(56)$ & $0.7339(68)$ & $114.6(2.8)$ & $37.69(54)$ \cr
          & $90$ & $0.6691(50)$ & $0.7347(60)$ & $125.8(2.1)$ & $40.94(48)$ \cr
          & $100$ & $0.6734(46)$ & $0.7380(55)$ & $137.6(2.8)$ & $42.78(50)$ \cr
          & $110$ & $0.6728(42)$ & $0.7358(50)$ & $152.4(5.4)$ & $47.97(83)$ \cr
          & $120$ & $0.6761(39)$ & $0.7407(46)$ & $167.2(3.2)$ & $49.21(50)$ \cr
\hline
\vspace{-2mm}
& & & & & \cr
Extrapolated & & $0.6884(54)$ & $0.760(14)$ \cr
\hline
%\vspace{-1mm}
%& & & & & \cr
%\multicolumn{3}{l}{$\dagger$ -- Comment} \\
\end{tabular}
\end{indented}
\label{t2}
\end{table}

The data in table \ref{t1} can be extrapolated to determine the limiting critical 
point(s) $\delta_c^{(m)}$ where the arms of the $2$-star escape.   Plotting
$\delta_c^{(m)}$ against $1/L$ shows a linear relationship, and so we bootstrap
the data using the linear model $\delta_L^{(m)} = \delta_c^{(m)} + b_0/L$ and 
a weighted least squares fit.   A bootstrap over $10{,}000$ iterations gives 
$\delta_c^{(1)} = 0.6884(54)$ (here, and in subsequent bootstrap estimates,  
the error bar is not a statistical standard deviation, but is a confidence interval 
which is twice the standard deviation of the dispersion or spread of the individual 
boot-strap estimates).  Thus, our best estimate of the location of the first arm to
escape is 
\begin{equation}
\delta_c^{(1)} = 0.6884 \pm 0.0054,
\quad\hbox{for $2$-stars with $w=3$ in the square lattice}.
\end{equation}
Neglecting the data point at $L=10$ gives $0.6903(46)$, consistent with the
estimate above.

Extrapolating $\delta_c^{(2)}$ is less effective, since there are fewer data points.
Plotting $\delta_L^{(2)}$ against $1/L$ shows that the data are approximately
linear when $L\geq 70$.  Bootstrapping a linear model including all the data gives
$\delta_c^{(2)} = 0.760(14)$, while neglecting the data points up to $L=60$ gives
$\delta_c^{(2)} = 0.7527(68)$.    This last estimate is within the error
bar of the first, and we therefore take as our best estimate
\begin{equation}
\delta_c^{(2)} = 0.760 \pm 0.014,
\quad\hbox{for $2$-stars with $w=3$ in the square lattice}.
\label{eqn6}
\end{equation}

The heights of the peaks in figure \ref{F4} should increase with $L$ according to
\begin{equation}
h_c^{(m)} (L) \sim C_m\,L^{2\phi_m-1}
\end{equation}
where $\phi_m$ is a finite size crossover exponent.  Estimating $\phi_m$
gives an effective value of $\phi_m$, and these are functions of $L$.  By taking
a ratio and then logarithms, finite size estimates of $\phi_m$ can be determined: 
\begin{equation}
2\phi_m-1 = \frac{\log( h_c^{(m)}(L_1)/h_c^{(m)}(L_2) )}{\log(L_1/L_2)} .
\label{eqn8}
\end{equation}
For pairs $(L_1,L_2)$ estimates of $2\phi_m-1$ are obtained, and bootstrapping
these over $10{,}000$ iterations give $\phi_1=0.97(12)$ and $\phi_2=1.16\pm0.82$
(where the error bars again are twice the standard deviation of the dispersion 
of the boot-strap estimates).  These results are not inconsistent with the 
escape transitions being first order, as suggested by the sharp rises in 
the pressure $P(\delta)$ in the right panel of figure \ref{F3}.

\subsection{Mean total span}
The \textit{total span} of an $f$-star is the sum of the spans of its arms, 
namely $\sigma=\sigma_1+\sigma_2+\cdots+\sigma_f$.  Recall from equation 
\Ref{eqn1} that we can count the number of stars with total span $\sigma$ 
by a restricted sum over sequences $[\sigma_i]$:
\begin{equation}
S_n^{(f)} (\sigma;w,L) 
 = {\sum_{[\sigma_i]}}^\prime  S^{(f)}_n([\sigma_i];w,L), 
\end{equation}
where the primed summation is over all $[\sigma_i]$ such that
$\sum_i \sigma_i = \sigma$.  
Then the \textit{mean total span} is defined by
\begin{equation}
\sigma^{(mean)}_n = \frac{\sum_\sigma \sigma\, 
S_n^{(f)}(\sigma;w,L)}{S_n^{(f)}(w,L)} .
\end{equation}

There are several bounds on the mean total span.  The first is a lower bound 
using the minimum span, obtained by filling the pore from the left.  If 
$n\leq (w{+}1)^dL$, then
\begin{equation}
\sigma^{(mean)}_n \geq \lfl n/(w+1)^d \rfl .
\end{equation}
It trivially the case that $\sigma^{(mean)}_n \leq n$. These bounds show that, 
for large $L$ (say $L > \lcl n/(w{+}1)^d \rcl$,  the mean total span, and the 
mean spans of individual arms,  are $\Theta(n)$ functions.

\begin{figure}
  \includegraphics[width=0.5\textwidth,height=0.3\textheight]{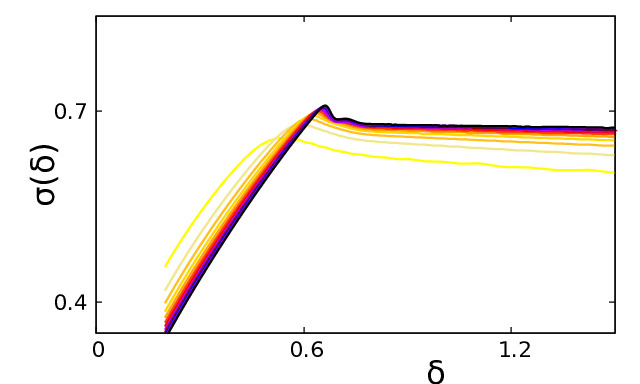}
  \includegraphics[width=0.5\textwidth,height=0.3\textheight]{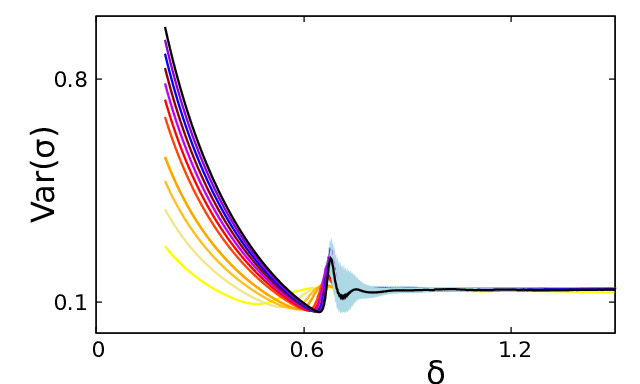}
  \caption{(Left) The span per unit length as a function of the relative pore
  depth $\delta$.  For large $\delta$ the data approach a constant as $L$ increases,
  consistent with equation (\ref{eqn12}).  If $\delta$ is small, then part of the
  walk escapes, and the total span per unit length decreases with decreasing
  $\delta$. The curves have maxima at the critical points, where the arms of
  the star escape.  (Right) The variances of the curves in the left panel.  The variances
  have local maxima at the critical points, and undulates as $\delta$ decreases towards
  the critical point(s).  The increase in the variance as $\delta$ decreases towards zero
  is expected, since the dispersion of the arms in bulk will be greater than the
  dispersion when the arms are retracted inside the pore. The local maxima in these
  curves correspond to critical points in the model.  The colours of the curves 
  increases to  black as $L$ increases in $\{10,20,\ldots,120\}$.  
  The shaded envelope in the right panel is the measured standard deviation about 
  the plotted curve for $L=120$.}
  \label{F5}
\end{figure}

It is expected that the part of an escaped arm outside the pore will have span 
given by $O(n^\nu)$ where $\nu$ is the metric exponent of the self-avoiding walk.  
The part inside the pore will grow as $O(\delta n / f) \sim O(L)$.  These considerations 
show that the mean total span \textit{per unit length} should approach a 
positive constant if $\delta$ is sufficiently large so that none of the arms 
of the star have escaped. On the other hand, if $\delta$ is small and the arms
have escaped from the pore, then the total span per unit length will be a 
decreasing function of decreasing $\delta$.  This is seen, for square lattice 
$2$-stars, in the left panel of figure \ref{F5}.  With increasing $L$, the total 
mean span per unit length accumulates on a horizontal locus for $\delta$ large, 
but for small values of $\delta$ the arms of the star escape and the curves 
accumulate on a curve which is $O(\delta)$.  This is consistent with 
\begin{equation}
\limsup_{n\to\infty} \sfrac{1}{n} \sigma_n^{(mean)}%\quad
\sim
\cases{
  K>0, & \hbox{if $\delta > \delta_c$}; \\
  O(\delta), & \hbox{if $\delta < \delta_c$}.
}
\label{eqn12}
\end{equation}

Since the mean total span has distinctly different behaviour in the escaped 
and retracted phases, we can use this quantity for an alternative estimate 
of the location of the escape transitions $\delta_c$.  Indeed, the data for 
mean total spans in the left panel of figure \ref{F5} shows a global maximum, 
indicating the dominant escape transition, and, for larger $L$, may have additional 
local maxima indicating secondary escape transitions.  For small $L$ there are 
also undulations for $\delta > \delta_c$.  The source of these are unclear, but may 
be due to parity effects in our data.  Label the peaks, as before, from the left to 
be located at $\delta_s^{(j)}$ for $j=1,2$ where $j=1$ is the dominant peak 
and $j=2$ the secondary peak.  In table \ref{t3} the $\delta_s^{(j)}$ are listed, 
while the heights of the local maxima are denoted by $\Sigma_L^{(j)}$.

\begin{table}[h!]
\caption{Statistics of the total span of $2$-stars in a pore of width $3$ and depth $L$}
\begin{indented}
\lineup
\item[]
\begin{tabular}{@{}*{3}{l}{l}llllll}
%\hline          
Model & $L$ & $\delta_s^{(1)}$ & $\delta_s^{(2)}$ & $\Sigma_L^{(1)}$ & $\Sigma_L^{(2)}$ \cr
\hline
\vspace{-2mm}
& & & & & \cr
\multirow{2}{*}{$2$-Stars} 
          & $10$ & $0.548(62)$ & $-$ & $0.6573(38)$ & $-$ \cr
\multirow{2}{*}{$d=2$} 
          & $20$ & $0.592(36)$ & $-$ & $0.6784(25)$ & $-$ \cr
\multirow{2}{*}{$w=3$} 
          & $30$ & $0.615(26)$ & $-$ & $0.6877(21)$ & $-$ \cr
          & $40$ & $0.627(20)$ & $-$ & $0.6932(18)$ & $-$ \cr
          & $50$ & $0.635(17)$ & $-$ & $0.6969(17)$ & $-$ \cr
          & $60$ & $0.640(14)$ & $-$ & $0.6996(16)$ & $-$ \cr
          & $70$ & $0.647(13)$ & $-$ & $0.7020(15)$ & $-$ \cr
          & $80$ & $0.649(11)$ & $0.711(13)$ & $0.7029(14)$ & $0.6821(1)$ \cr
          & $90$ & $0.6557(96)$ & $0.710(12)$ & $0.7053(14)$ & $0.6867(2)$ \cr
         & $100$ & $0.6568(87)$ & $0.718(11)$ & $0.7060(12)$ & $0.6855(4)$ \cr
         & $110$ & $0.6597(80)$ & $0.7225(96)$ & $0.7072(12)$ & $0.6862(7)$ \cr
         & $120$ & $0.6621(74)$ & $0.7218(87)$ & $0.7081(11)$ & $0.6877(3)$ \cr
\hline
\vspace{-2mm}
& & & & & \cr
Extrapolated & & $0.7097(32)$ & $0.7856(93)$ & $0.7123(36)$ & $0.700(12)$  \cr
\hline
\end{tabular}
\end{indented}
\label{t3}
\end{table}

The data in table \ref{t3} can again be extrapolated to find limiting estimates
of the local maxima and heights of the curves. Numerical experimentation suggest 
the model $a_0 + b_0/L$ and boot-strapping using a weighted least squares fit give
the estimate of the dominant transition at $\delta_s^{(1)} = 0.7097\pm0.0032$ 
with limiting height per unit length of the peak $\Sigma^{(j)}=0.7123\pm 0.0036$. 
In the case of $\delta_s^{(2)}$ there are only 5 data points, and this makes 
extrapolation more uncertain.  Using the same approach as above gives 
$0.786\pm0.020$, an estimate slightly above the result in equation \Ref{eqn6}.
Since there are $7$ data points available for the estimate in equation \Ref{eqn6},
only $5$ from the data in table \ref{t3}, we take equation \Ref{eqn6} as our best
estimate of the location of the escape of the second arm.

The data on the heights of the peaks in figure \ref{F5} listed in table \ref{t3}
suggest that the peaks may each be approaching a constant with increasing $L$.
Putting $\Sigma^{(i)} = \lim_{L\to\infty} S_L^{(i)}$ as the limiting height of the peaks,
extrapolating the data in table \ref{t3} gives $\Sigma^{(1)} = 0.7123(36)$ for the 
dominant peak.  There are less data to extrapolate effectively to the secondary
peak, but boot-strapping from only five data points give $\Sigma^{(2)} = 0.700(12)$.
In both cases the heights of the peaks appear to approach a constant as $L\to\infty$.

\begin{figure}
\centering
\includegraphics[width=0.45\textwidth,height=0.3\textheight]{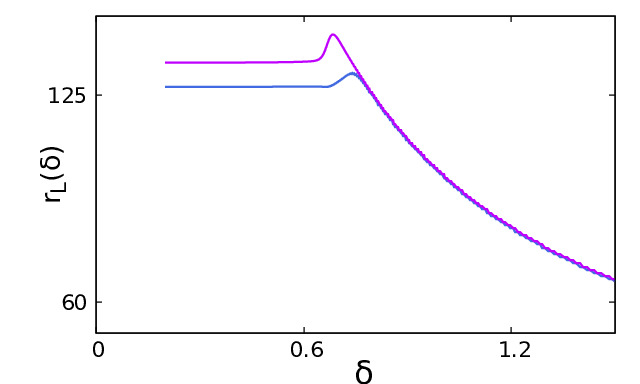}
\includegraphics[width=0.45\textwidth,height=0.3\textheight]{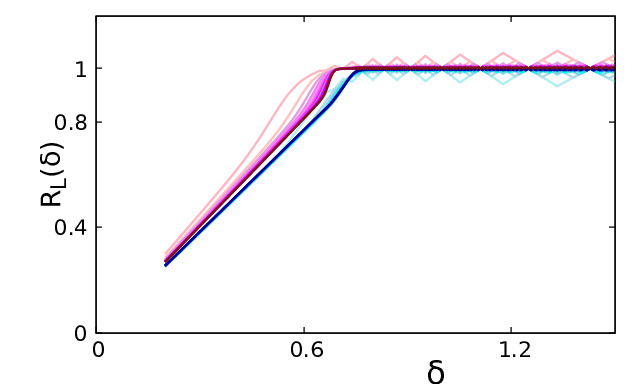}
\caption{(Left) The length of arms inside the pore for $L=100$ for $2$-stars
in the square lattice in a pore of width $w=3$, plotted as a function of 
the relative pore depth $\delta = Lf/n = 200/n$.  Decreasing $\delta$ from the right
(by increasing the total length $n\in [1,1000]$ of the star) increases the length of 
the arms inside the pore $r_L^{(i)}(n)$ until the escape transition occurs, whereafter
a constant length of the arms are inside the pore even as the lengths of the
escaped arms continue to  increase while $\delta$ decreases. The escape transitions
are signalled by small peaks in the curves, one for each escape. 
(Right) The fraction of arms inside the pore as a function of the relative pore depth 
$\delta$ for square lattice $2$-stars in a pore of width $w=3$.  For large $\delta$ 
the arms are retracted in the pore, and $R_L^{(i)} = 1$ (the periodic jumps in 
the data are due to lattice parity effects).  Decreasing $\delta$ causes the arms 
to escape.  In the escaped phase (small $\delta$) only a part of the arm remains
inside the pore and thus $R_L^{(i)}$ decreases.  There are two escape transitions 
in this model, one for the escape of each of the two arms.  Data were collected 
for $L$ in steps of $10$ from $10$ to $120$.}
\label{F6}
\end{figure}

\subsection{Fractions of arms within the pore}
If an arm escapes from the pore it has a last node inside the pore before it
exits for the first time.  This vertex is the \textit{exit node} of the arm.  An example 
of an exit node is shown in figure \ref{F2}.  If the arm does not escape, then place 
the exit node, by default, at the last vertex of the arm.  The chemical distance 
(along the arm) from the central node of the star to the exit node, or to the
endpoint of the arm, whichever is first, is in the interval $[0,\lcl n/f \rcl]$.   
We denote the mean of this distance, namely the \textit{mean chemical distance},
by $r_L^{(i)}(n)$ (and this is a function of $(f,w,L,n)$). 

If $\lfl n/f \rfl \leq r_L^{(i)}(n) \leq \lcl n/f \rcl$ then the $i$-th arm of the 
star is (said to be) retracted inside the pore, and if $r_L^{(i)}(n)  < \lfl n/f \rfl$ 
then it is said to have escaped from the pore.  In the left panel of figure \ref{F6}
the mean chemical distance $r_L^{(i)}(n)$ along the two arms of 
a $2$-star of length $n\in[0,1000]$ are plotted for a pore of depth 
$L=100$ as a function of relative pore depth $\delta_n = 200/n$.
Once the two arms have escaped, then the mean chemical distance for 
each are relatively constant.  In this particular example the mean chemical
distance of the primary escaped arm to its exit node is 
$r_{100}^{(1)}\approx 135.2$ while the same distance for the secondary 
escaped arm is $r_{100}^{(2)}\approx 127.5$. We notice a small overshoot
in both the arms as the escape transitions occur in the left panel of 
figure \ref{F6}.  The pore accommodates a slightly longer mean chemical 
distance near the critical point for each arm, but once an arm escapes then 
it tightens inside the pore, reducing its mean chemical distance.

 In the limit as $n\to\infty$ define the fraction of the $i$-th arm in
the pore of length $L=\lfl\delta n \rfl$ by
\begin{equation}
\rho^{(i)}(\delta) = 
\limsup_{n\to\infty} \frac{r_{\lfl \delta n\rfl}^{(i)}(n)}{\lfl n/f \rfl} .
\end{equation}
If $\delta>1$, then the arm cannot escape and it is wholly contained
in the pore.  In this case $r_{\lfl \delta n\rfl}^{(i)}(n)$ is almost equal
to the length of the arm, and $\rho^{(i)}(\delta) = 1$.  On the 
other hand, for small values of $\delta$, the arm escapes
and $r_{\lfl \delta n\rfl}^{(i)}(n)$ is strictly less than $\lcl n/f \rcl$.
In this event one expects $\delta \leq \rho^{(i)}(\delta) < 1$.

Finite size approximations to $\rho^{(i)}(\delta)$ are given by
$R_L^{(i)}(\delta) = r_{\lfl \delta n\rfl}^{(i)}(n) / \lfl n/f \rfl$.  If $n$ is small
(large $\delta$) in the left panel of figure \ref{F6} then the arms of the
star are retracted inside the pore, and thus $R_L^{(i)}\approx 1$.  Increasing 
$n$ reduces $\delta$, and once parts of the arms escape, $R_L^{(i)}$ 
decreases since the fraction of the arms inside the pore is decreasing.  
This is seen in the right panel of figure \ref{F6}.

\begin{figure}
\includegraphics[width=0.5\textwidth,height=0.3\textheight]{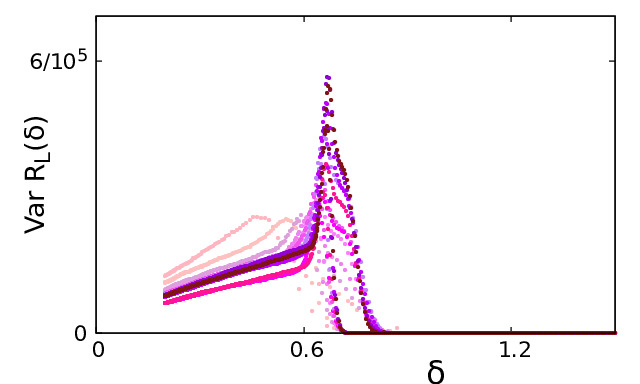}
\includegraphics[width=0.5\textwidth,height=0.3\textheight]{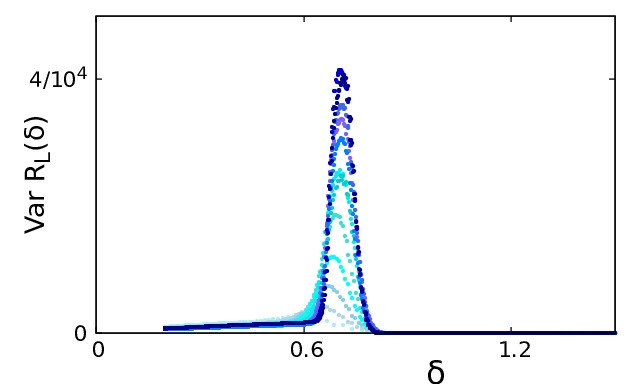}
\caption{Variances of the fraction of arms inside the pore for square lattice 
$2$-stars in a pore of width $w=3$.  
(Left)  The variance of the fraction of the arm inside the pore corresponding
to the primary escape transition.  The data points are asymmetric around 
the critical point and peak sharply at the critical point.
(Right) The variance corresponding to the secondary escape transition.  Notice 
the large difference in scale on the vertical axes when comparing the two plots.  
The variance of the primary escape transition exhibits considerable oscillations 
and asymmetry around the critical point, partly due to lattice parity effects.  
At the primary transition on the left the secondary arm is already in the escaped
phase, and its experiences a conformational rearrangement when the primary 
arm also escapes and vacates space inside the pore.  This may explain the 
the asymmetric distribution of data points around the peak on the left.  In contrast, 
the peak on the right corresponding to the secondary escape transition which occurs
when the primary arm is still retracted inside the pore.  When the secondary arm
escapes, it also vacates space inside the pore, but this is then explored by 
conformations of the retracted arm without having an impact on the fraction 
of the completely retracted primary arm inside the pore. Data were collected for 
$L$ in steps of $10$ from $10$ to $120$.}
\label{F7}
\end{figure}

\begin{table}[h!]
\caption{Fraction of arms of $2$-stars inside a pore of width $3$, depth $L$}
\begin{indented}
\lineup
\item[]
\begin{tabular}{@{}*{3}{l}{l}llll}
Model & $L$ & $\delta_c^{(1)}$ & $\delta_c^{(2)}$ & 
  $\delta_v^{(2)}$ & $\delta_v^{(2)}$ & $h_v^{(1)}\times 10^5$ & $h_v^{(2)}\times 10^4$   \cr
\hline
\vspace{-2mm}
& & & & & \cr
\multirow{2}{*}{$2$-Stars} 
          & $10$ & $0.556(31)$ & $0.667(45)$ & $0.465(22)$ & $0.588(35)$ & $2.557(52)$ & $0.2622(98)$ \cr
\multirow{2}{*}{$d=2$} 
          & $20$ & $0.625(20)$ & $0.714(26)$ & $0.548(16)$ & $0.645(21)$ & $2.517(37)$ & $0.431(11)$ \cr
\multirow{2}{*}{$w=3$} 
          & $30$ & $0.645(14)$ & $0.714(18)$ & $0.594(12)$ & $0.667(15)$ & $2.60(13)$ & $0.7495(80)$ \cr
          & $40$ & $0.656(11)$ & $0.727(14)$ & $0.6202(97)$ & $0.684(12)$ & $2.57(20)$ & $1.2075(98)$ \cr
          & $50$ & $0.6667(89)$ & $0.735(11)$ & $0.6369(82)$ & $0.6944(97)$ & $2.94(43)$ & $1.865(71)$ \cr
          & $60$ & $0.6742(76)$ & $0.7317(89)$ & $0.6486(71)$ & $0.7059(84)$ & $3.47(60)$ & $2.569(65)$ \cr
          & $70$ & $0.6796(66)$ & $0.7368(78)$ & $0.6573(62)$ & $0.7071(72)$ & $2.97(70)$ & $2.482(82)$ \cr
          & $80$ & $0.6808(58)$ & $0.7407(69)$ & $0.6584(55)$ & $0.7080(63)$ & $4.34(88)$ & $3.369(69)$ \cr
          & $90$ & $0.6818(52)$ & $0.7438(62)$ & $0.6642(50)$ & $0.7087(56)$ & $3.73(83)$ & $3.069(24)$ \cr
          & $100$ & $0.6849(47)$ & $0.7407(55)$ & $0.6645(45)$ & $0.7143(52)$ & $5.08(78)$ & $3.593(34)$ \cr
          & $110$ & $0.6875(43)$ & $0.7432(51)$ & $0.6687(41)$ & $0.7074(46)$ & $5.66(95)$ & $4.149(38)$ \cr
          & $120$ & $0.6897(40)$ & $0.7407(46)$ & $0.6723(38)$ & $0.7143(43)$ & $5.4(12)$ & $4.067(96)$ \cr
\hline
\vspace{-2mm}
& & & & & \cr
Extrapolated & & $0.7015(38)$ & $0.7500(40)$ & $0.6935(66)$ & $0.7264(62)$  \cr
\hline
\end{tabular}
\end{indented}
\label{t4}
\end{table}

The escapes of the individual arms are also seen in the variances of 
$R_L^{(i)}(\delta)$.  These are plotted in figure \ref{F7}.  The left
panel shows the variance associated with the primary escape transition.  This 
is the dominant escape transition.  The secondary escape corresponding to the escape of
the second arm corresponds to a larger variance, shown in the right panel.
Notice in particular the difference in scale on the vertical axes of the two 
plots. The variance associated with the primary transition is asymmetric and has 
significant parity oscillations, showing that there is a conformational rearrangement 
of the star involving both arms at the primary escape transition.  One may guess
that the mechanism underlying the asymmetry and parity effects seen in the
left panel occurs because the escaping arm vacates space in the pore, impacting 
the conformational space available to the (still retracted) other arm.

The locations $\delta_c^{(i)}$ of the peaks in figure \ref{F6}, of the
location of the peaks $\delta_v^{(i)}$ in the variances in figure \ref{F7},
with the associated heights of these peaks $h_v^{(i)}$, are shown in table \ref{t4}.
The estimates of $\delta_c^{(i)}$ and $\delta_v^{(i)}$ can again be extrapolated
using a boot-strap as before, giving the results in the last row of table
\ref{t4}.  These estimates are consistent with our best estimates in 
table \ref{t1} in the sense that doubling error bars in table \ref{t1} intersects
the confidence intervals in table \ref{t4}.

\section{Cubic lattice stars}
\label{sec:CubicStars}

\begin{figure}
\includegraphics[width=0.475\textwidth,height=0.35\textheight]{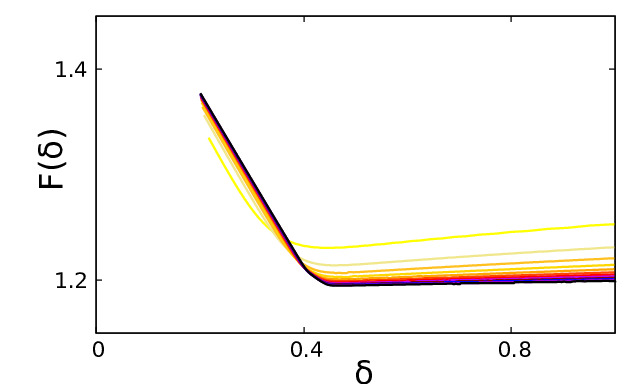}
\includegraphics[width=0.475\textwidth,height=0.35\textheight]{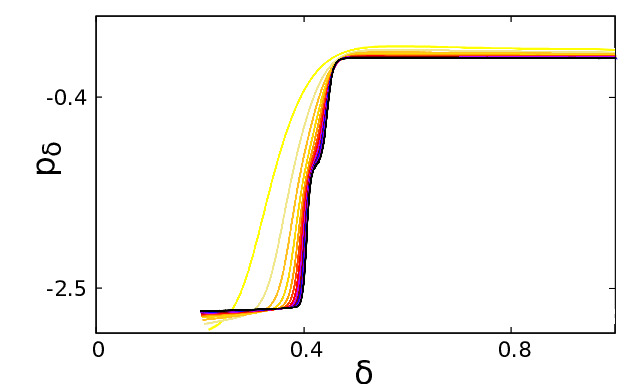}
\caption{(Left)  The free energy of cubic lattice $2$-stars with arms escaping
from a $3\times 3\times L$ pore.  The hue of the colours increases from yellow to 
black as $L$ increases from $10$ to $120$.  Stars of total length up to $n=10 L$
were sampled and the relative pore depth is given by $\delta=L/n$.  
(Right) The pressure (the derivative of the free energy to $\delta$) in the pore.
There is a sharp decrease (from a large negative pressure to a smaller negative 
pressure) of the pressure as $\delta$ increases through the escape transitions.
Closer inspection of the data reveals a two-step decrease, one for each of the 
two escape transitions in this model.}
\label{F8}
\end{figure}

\begin{figure}[t!]
\includegraphics[width=0.475\textwidth,height=0.35\textheight]{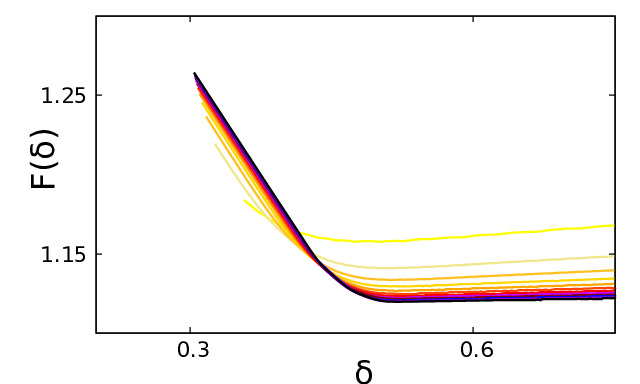}
\includegraphics[width=0.475\textwidth,height=0.35\textheight]{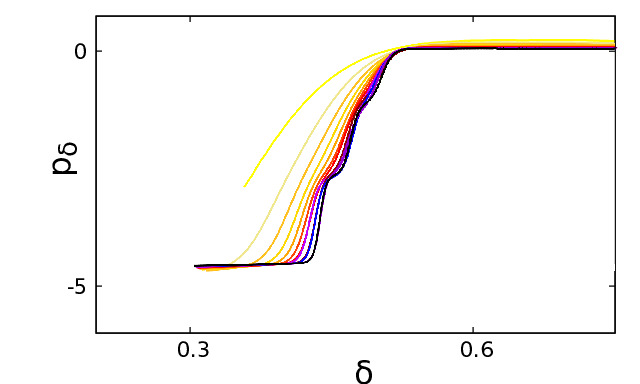}
\caption{(Left)  The free energy of cubic lattice $3$-stars with arms escaping
from a $3\times 3\times L$ pore.  The hue of the colours increases from yellow to 
black as $L$ increases from $10$ to $120$.  Stars of total length up to $n=10 L$
were sampled and the relative pore depth is given by $\delta=L/n$.  
(Right) The pressure (the derivative of the free energy to $\delta$) in the pore
changes (from a large negative pressure to a smaller negative pressure)
as the relative pore depth $\delta$ increases through the escape transitions.
Closer inspection of the data reveals that the pressure changes in three 
steps with increasing relative pore depth $\delta$, each change corresponding 
to an escape transition of one of the three arms of the star.}
\label{F9}
\end{figure}

We have also collected data on cubic lattice stars.  In this case the model is 
again similar to the model in figure \ref{F2}, but the pore has dimensions
$w\times w\times L$ (that is, the bottom of the pore is a square of side-length $w$,
and the pore has depth $L$).  The central node of the star is grafted at the vertex
$(0,\lfl w/2 \rfl, \lfl w/2 \rfl)$.  The methods discussed in section \ref{sec:NumericalResults} 
are again used to estimate the locations of the escape transitions.
The free energy and pressure of $2$-stars grafted in $3\times 3\times L$ pore
are plotted in figure \ref{F8}, and for $3$-stars in figure \ref{F9}.  The
pressure gradients for both models are plotted in figure \ref{F10}.

\begin{table}[h!]
\caption{Escape transitions of cubic lattice $2$-stars in a pore of width $3$ and depth $L$}
\begin{indented}
\lineup
\item[]
\begin{tabular}{@{}*{3}{l}{l}llll}
%\hline          
Dimension & $L$ & $\delta_c^{(1)}$ & $\delta_c^{(2)}$ & $h_c^{(1)}$ & $h_c^{(2)}$  \cr
\hline
\vspace{-2mm}
& & & & & \cr
\multirow{2}{*}{$2$-stars} 
          & $10$ & $0.323(11)$ & $-$ & $75.63(64)$ & $-$ \cr
\multirow{2}{*}{$d=3$} 
          & $20$ & $0.3636(67)$ & $-$ & $97.82(90)$ & $-$ \cr
\multirow{2}{*}{$w=3$} 
          & $30$ & $0.3750(47)$ & $-$ & $119.23(91)$ & $-$ \cr
          & $40$ & $0.3846(37)$ & $0.4211(45)$ & $141.4(11)$ & $81.61(56)$ \cr
          & $50$ & $0.3906(31)$ & $0.4310(38)$ & $164.9(13)$ & $90.11(70)$ \cr
          & $60$ & $0.3941(26)$ & $0.4364(32)$ & $190.3(17)$ & $101.33(53)$ \cr
          & $70$ & $0.3977(23)$ & $0.4389(28)$ & $216.9(17)$ & $113.92(46)$ \cr
          & $80$ & $0.4000(20)$ & $0.4401(25)$ & $243.4(23)$ & $127.12(69)$ \cr
          & $90$ & $0.4018(18)$ & $0.4434(22)$ & $270.1(19)$ & $141.9(13)$ \cr
        & $100$ & $0.4040(17)$ & $0.4435(20)$ & $298.2(25)$ & $154.31(88)$ \cr
        & $110$ & $0.4052(15)$ & $0.4444(18)$ & $327.4(27)$ & $166.59(52)$ \cr
        & $120$ & $0.4054(14)$ & $0.4453(17)$ & $346.1(52)$ & $182.3(34)$ \cr
\hline
\vspace{-2mm}
& & & & & \cr
Extrapolated & & $0.4140(33)$ & $0.4565(31)$ \cr
\hline
\end{tabular}
\end{indented}
\label{t8}
\end{table}

\subsection{2-stars}
\begin{figure}[t!]
\includegraphics[width=0.475\textwidth,height=0.35\textheight]{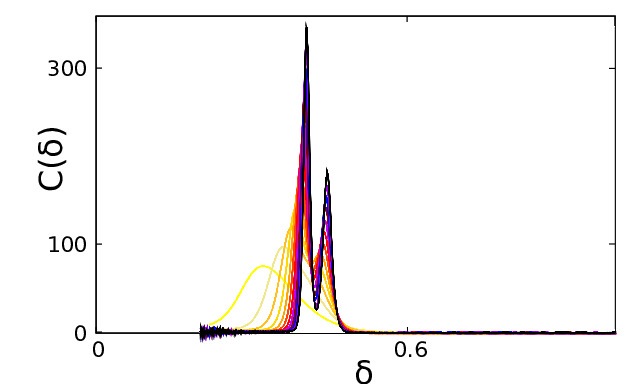}
\includegraphics[width=0.475\textwidth,height=0.35\textheight]{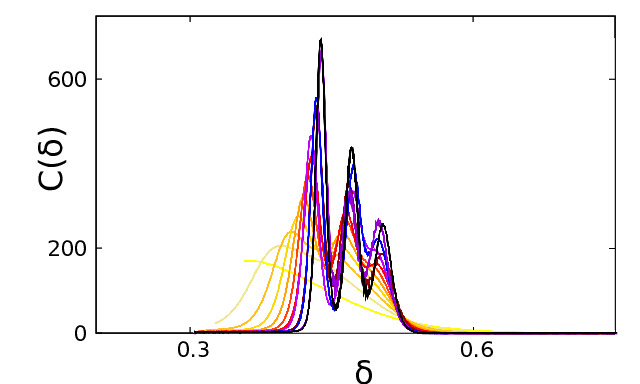}
\caption{(Left) The pressure gradients of escape transitions of $2$-stars 
from a $3\times 3\times L$ pore in the cubic lattice.  The pressure gradient 
shows two clearly developed peaks which become prominent when $L>40$.  
The dominant peak corresponds to the primary escape of an arm from the pore, 
and the secondary peak corresponds to the escape of the remaining second arm.  
(Right) The pressure gradient of escape transitions of $3$-stars from a $3\times 3\times L$
pore in the cubic lattice. The left-most dominant peak corresponds to the
primary escape transition in the model, while the remaining two peaks 
correspond to secondary transitions of the remaining arms.  The data show
show three clearly developed peaks when $L>70$.  The colours of the curves 
increase to  black as $L$ increases in $\{10,20,\ldots,120\}$.}
\label{F10}
\end{figure}

The free energy, pressure and pressure gradient of an escaping cubic lattice 
$2$-star are shown in figures \ref{F8} and \ref{F10}.  The characteristics
of these plots are similar to the square lattice results in figures \ref{F3} and \ref{F4},
with two escape transitions developing in the model as $L$ is increased from
$L=10$ to $L=120$ in steps of $10$. Extrapolating the critical values of 
$\delta_c^{(i)}$ gives the estimates
\begin{equation}
\delta_c^{(1)} = 0.4140 \pm 0.0033,\quad
\delta_c^{(2)} = 0.4565 \pm 0.0031,
\end{equation}
for the locations of the critical points of the two escape transitions in this model.

The heights of the peaks can also be examined using equation \Ref{eqn8} to 
determine a finite size crossover exponent associated with the peaks in the
pressure gradient.  This gives $\phi_1=1.04 \pm 0.12$, and $\phi_2=1.21 \pm 0.09$,
where the error bars are twice the standard deviations of the dispersion of the 
boot-strap estimates.

\begin{table}[h!]
\caption{Escape transitions of cubic lattice $3$-stars in a pore of width $3$ and depth $L$}
\begin{indented}
\lineup
\item[]
\begin{tabular}{@{}*{3}{l}{l}llllll}
%\hline          
Dimension & $L$ & $\delta_c^{(1)}$  & $\delta_c^{(2)}$ & $\delta_c^{(3)}$ 
                  & $h_c^{(1)}$ & $h_c^{(2)}$   & $h_c^{(3)}$  \cr
\hline
\vspace{-2mm}
& & & & & \cr
\multirow{2}{*}{$3$-stars} 
         & $10$ & $0.361(44)$   & $-$ & $-$ & $171.79(73)$ & $-$ & $-$ \cr
\multirow{2}{*}{$d=3$} 
         & $20$ & $0.4000(54)$ & $-$ & $-$ & $206.39(97)$ & $-$& $-$ \cr
\multirow{2}{*}{$w=3$} 
         & $30$ & $0.4110(38)$ & $-$ & $-$ & $239.6(1.2)$ & $-$& $-$ \cr
         & $40$ & $0.4167(29)$ & $0.4598(36)$ & $-$ & $275.6(1.4)$ & $214.8(1.1)$& $-$ \cr
         & $50$ & $0.4225(24)$ & $0.4630(29)$ & $-$ & $322.6(1.2)$ & $238.1(1.6)$& $-$ \cr
         & $60$ & $0.4265(21)$ & $0.4675(25)$ & $-$ & $372.1(2.4)$ & $265.5(1.0)$& $-$ \cr
         & $70$ & $0.4277(18)$ & $0.4688(21)$ & $0.4988(24)$ & $437.8(6.4)$ & $275.8(8.2)$& $169.5(2.7)$ \cr
         & $80$ & $0.4293(16)$ & $0.4724(17)$ & $0.4969(21)$ & $464.8(1.6)$ & $336.6(5.4)$& $197.7(1.7)$ \cr
         & $90$ & $0.4355(15)$ & $0.4704(17)$ & $0.5056(19)$ & $536.3(4.6)$ & $354.5(6.3)$& $186.9(1.6)$ \cr
        & $100$ & $0.4354(13)$ & $0.4739(15)$ & $0.5009(17)$ & $556.3(5.7)$ & $395.1(6.3)$& $223.5(1.5)$ \cr
        & $110$ & $0.4400(12)$ & $0.4701(14)$ & $0.5008(16)$ & $663.2(11.5)$ & $337.3(7.8)$& $267.2(7.6)$ \cr
        & $120$ & $0.4396(11)$ & $0.4718(13)$ & $0.5056(15)$ & $693.2(5.6)$ & $437.3(2.1)$& $256.9(3.0)$ \cr
\hline
\vspace{-2mm}
& & & & & \cr
Extrapolated & & $0.4481(64)$ & $0.4782(53)$ & $0.5118(88)$ \cr
\hline
%\vspace{-1mm}
%& & & & & \cr
%\multicolumn{3}{l}{$\dagger$ -- Comment} \\
\end{tabular}
\end{indented}
\label{t11}
\end{table}

\subsection{3-stars}
Data were also collected using a model of $3$-stars with arms escaping from 
a $w\times w\times L$ pore in the cubic lattice.   The free energy, pressure and 
pressure gradient are plotted for this model in figures \ref{F9} and \ref{F10}.  
The free energy curves have shape similar to those seen in figure \ref{F8}, 
but the pressure curves show additional structure suggesting step-wise 
increases with $\delta$ when transitions corresponding to escaping arms 
occur.  Indeed, the pressure gradient in the right-most panel in figure 
\ref{F9} shows three well-separated peaks suggesting that the arms 
of the stars escape one at a time, each time leading to a conformational 
rearrangement of the parts of the star remaining in the pore.

Similar to the case for $2$-stars, the data in figure \ref{F9} were examined and
the locations and heights of the peaks in the pressure gradient determined.  
For $L\leq 30$ only one peak was seen in the pressure gradient, which
became two peaks for $40\leq L\leq 60$ resolving into three well separated
peaks when $L\geq 70$, as seen in the right panel of figure \ref{F9}.  Our 
estimates of the peak locations in our data are listed in table \ref{t11}, and 
are extrapolated to 
\begin{equation}
\delta_c^{(1)} = 0.4481 \pm 0.0064,\quad
\delta_c^{(2)} = 0.4782 \pm 0.0053,\quad
\delta_c^{(3)} = 0.5118 \pm 0.0088.
\end{equation}
The growth in the height of the dominant peak in figure \ref{F9} can also be
examined to determine the estimate $\phi=1.05\pm 0.17$ for the finite size
crossover exponent, consistent with a first order transition.  The increases in the
heights of the secondary peaks are uneven, and we did not manage to extract
consistent estimates of the crossover exponents associated with those transitions.

\begin{figure}[h!]
  \includegraphics[width=\textwidth]{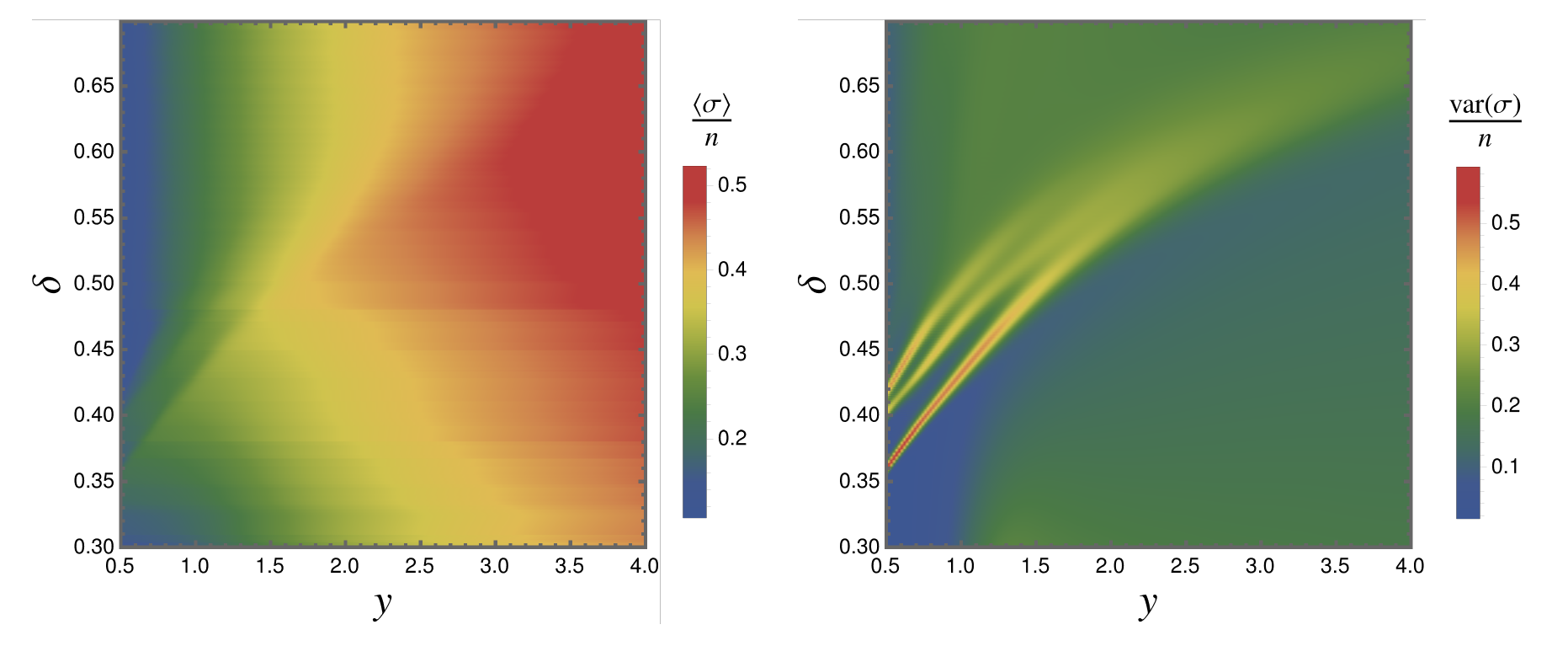}
  \caption{Phase diagrams for 3-stars in pores in the cubic lattice.   (Left) The left panel
  show the average of the total span $\sigma$ as a function of the relative pore
  depth $\delta$ (the vertical axis), and the fugacity $y$ conjugate to a pulling force
  along the pore axis on the arms of the star.  Large values of the total span are
  shown in red areas in the plot, while blue areas correspond to small values, with
  yellows and greens in between.  For small $y$ and large $\delta$ the arms are
  retracted and the span is small, while decreasing $\delta$ for fixed $y$ takes
  the model through escape transitions to phases with larger total span, even 
  in the regime where $y<1$ (and the force pushes the arms into the pore).
  (Right) The variance of the total span plotted instead.  These diagram shows
  three prominent curves of peaks running from the lower left to the upper right 
  across the phase diagram.  The primary transition corresponds to the lowest of 
  these transitions, and the presence of all three transitions for the range of $y$
  on the horizontal axis shows that the escape transitions persist over the 
  entire range of $y$ considered here, even if the height of the peaks decreases
  with increasing $y$.  Pore depth here is fixed at $L = 100$ in a 
  $3\times 3\times L$ pore with stars were simulated up to total length 
  $n=1000$.}
  \label{fig:CubicPhaseDiagram}
\end{figure}
% ================================================

% ==================================================
\section{Weighted total span}
\label{sec:WeightedSpan}
Next we consider an extension to this model by introducing an activity $y$ that 
is conjugate to the total span.  That is, we extend equation~\Ref{eqn1} to 
\begin{equation}
  S^{(f)}_n(y, w, L) = \sum_{\sigma=0}^n S^{(f)}_n(\sigma; w, L) \, y^\sigma.
\label{eq:PartitionFunctionExtended}
\end{equation}
Other quantities such as the free energy, mean total span, etc.~derived 
from equation~\Ref{eq:PartitionFunctionExtended} also become functions of 
$y$ in addition to $w,L,f,n$.  Results in previous sections correspond to $y=1$.
This approach is analogous to similar models where a force $F = k T \log y$ (in lattice units)
may be applied to the endpoint of a self-avoiding walk 
\cite{Rensburg2016,Beaton2015} or a single arm of an $f$-star \cite{Bradly2019c}.
For $f$-stars it is computationally prohibitive to have separate forces for each 
arm, hence although the addition of $y$ is similar to these pulled models in 
that it introduces an energetic component to the model, it does not have a 
clear physical interpretation in our case.

This extended model has a larger phase diagram when $y$ is varied.
The phase diagram is represented by the mean 
$\langle\sigma\rangle/(n/f)$ and variance $\mathrm{var}(\sigma)/(n/f)$ 
of the total span, normalised by arm length, as a function of $y$ and $\delta$, 
shown in \fref{fig:CubicPhaseDiagram}.  These plots show data for the case 
with fixed $L=100$ and $w=3$, up to length $n=1000$.  Varying $y$ and 
$\delta$ produces some interesting features.  The original model is present 
at $y=1$ where there is a phase at large $\delta$ in which the arms are 
retracted, and the average total span is independent of $\delta$.  As $\delta$ 
decreases there are three phase transitions close together as each arm escapes 
from the pore.  Then for small $\delta$ the average total span decreases as 
$\delta$ decreases, in line with the behaviour seen in \fref{F5}.

As $y$ decreases from $y=1$ the locations of the escape transitions move to smaller 
$\delta$, indicated by the peaks in $\mathrm{var}(\sigma)$ tracing three 
nearly parallel lines in \fref{fig:CubicPhaseDiagram}.  At the same time, as 
$y$ decreases the peaks in the variance become taller and sharper, indicating 
that the escape transitions are harder to achieve. This is due to the need to 
overcome the additional energetic barrier due to the `force' pushing the arms 
into the pore.  For small $y$ and small 
$\delta$, the favoured configurations are those with small total span and 
long length confined to a narrow pore.  These configurations are difficult 
to sample accurately and so we do not have good enough data for this 
kind of configuration to reliably extend the phase diagram below $y \lesssim 0.5$.

As $y$ increases from $y=1$ then the escape transitions move to higher $\delta$.
This can be interpreted as stars needing an additional `force' to escape pores with 
larger (fixed) relative pore depth.  However, as $\delta$ increases further this cannot 
continue indefinitely since, if the relative pore depth is of order or larger than the 
arm length, then the force required to enable escape will diverge.  This is evident 
in \fref{fig:CubicPhaseDiagram} (right panel) where the line of escape transitions 
begins to deviate from the linear behaviour in the region near $y=1$.
Unfortunately, this is a weak effect and we cannot resolve whether 
the secondary escape transitions merge together or simply become too weak to identify.
This is further exacerbated by the fact that larger $\delta$ is realised 
by smaller length $n$ and so finite-size effects become more significant.

\section{Conclusion}
\label{Conclusion}

In this paper we explored the nature of the escape transition of a star polymer from
a pore using an (equilibrium) Monte Carlo analysis.  Our numerical data show that
in the lattice model of a star polymer grafted in a pore, the arms escape at distinct
critical points, with a distinct transition associated with each arm.  Moreover, the
data are consistent with earlier results that these escape transitions are first order 
in nature \cite{MYB99,HBKS07}.  In addition, we have determined the location
of the critical points in several lattice models by locating the peak in the pressure 
gradient in each of our models.  

\begin{figure}[h!]
\includegraphics[width=0.475\textwidth,height=0.30\textheight]{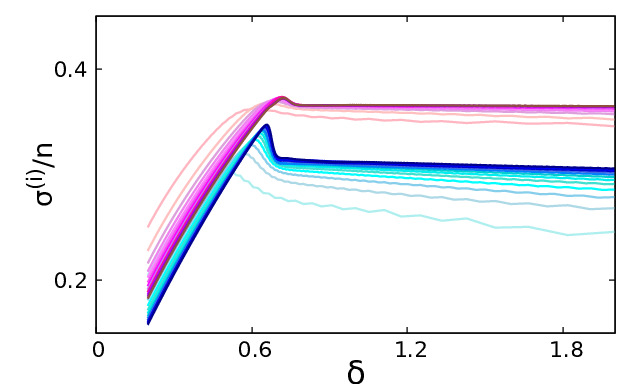}
\includegraphics[width=0.472\textwidth,height=0.30\textheight]{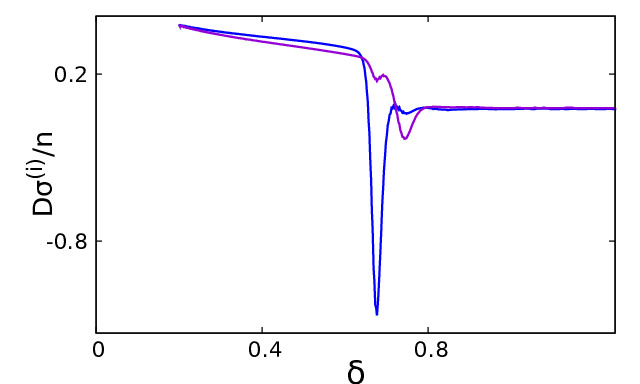}
\caption{(Left) The spans $\sigma^{(i)}(n)$ of the arm of a square lattice $2$-star 
in a pore of width $w=3$ as a fraction of the length $n$ of the star, plotted
as a function of the relative pore depth $\delta$.  The bottom family of curves
in hues of blue corresponds to the primary arm and escape transition of the
star, while the top family of curves to the secondary escape transition.  Reducing
$\delta$ from large values first takes the model through a secondary escape
transition (the smaller peak in the top family of curves) before the primary
transition seen as the larger peak in the bottom family of curves. (Right)  The 
derivative of $\sigma^{(i)}/n$ for $L=110$ and square lattice $2$-stars up 
to length $n=1100$.  The large and prominent dip in one curve corresponds 
to the primary transition on the left, while the secondary transition is seen in 
a smaller dip in the other curve.  Notice the smaller variations in both
curves corresponding to escape transitions in the other arms.}
\label{F13}
\end{figure}

Generally, the escape transitions of the arms in a lattice star are driven by
conformational rearrangements and interactions between the arms of the star 
inside the pore. The spans of the individual arms along the direction of the pore
can also be used as an indicator of the transitions, and in figure \ref{F13} the spans
$\sigma_n^{(i)}$ of the $i$-th arm of square lattice $2$-stars in a pore of width
$w=3$ are shown by plotting $\sigma_n^{(i)}/n$ against the relative pore depth
$\delta$.  The collection of curves for each arm are in different shades of blue, 
or of rose, and these increase in hue as the lengths of the star increases in steps 
of $10$ from $10$ to $120$.  The bottom family of curves (in shades of blue) 
corresponds to the primary escape transitions, and there is a prominent peak
in the data when the primary escape transition occurs.  This shows that the 
span of the primary arm is proportional to its length while it is retracted
inside the pore, but that it stretches along the direction of the pore as the
secondary arms escapes with decreasing $\delta$.  In other words, as
$\delta$ is decreased, and the secondary escape transition occurs (at the 
peak in the top family of curves in figure \ref{F13}), then the primary arm,
still retracted in the pore, occupies space vacated by the secondary arm 
inside the pore and expands its span inside the pore.  This is seen, for example,
in the right panel of figure \ref{F13}.  Here the primary escape is seen as a
dramatic dip in the derivative of $\sigma_n^{(1)}/n$ at the critical point, but
clearly visible in the other curve (the derivative of $\sigma_n^{(2)}/n$) is a small
variation as well.  This represents the response in the second arm as the first arm
undergoes its escape transition.  A similar, but smaller feature is seen in the other
curve corresponding the responds of the first arm with the second arms passes
through the secondary escape transition.

Finally, it should be noted that the escape transition of an arm of the star 
is a dynamical event, and on short time scales the escaping arm drives the rest 
of the star out of thermodynamic equilibrium.  Since polymers may have 
relatively long relaxation times, these non-equilibrium effects do eventually
die away as the star returns to thermal equilibrium, but they may have an
important impact on the nature of the escape transition, beyond the equilibrium
effects seen in our simulation.  Those effects are beyond the scope of the 
methods used in this paper.

\section*{Acknowledgement}
EJJvR acknowledges financial support from NSERC (Canada) in the form of 
a Discovery Grant RGPIN-2019-06303.   Data generated for this study
are available on reasonable request.   CJB acknowledges financial support 
from the Australian Research Council via its Discovery Projects 
scheme (DP230100674).

 %%%%%%%% Bibliography
\section*{References}
%\vspace{5mm}
\bibliographystyle{plain}
\bibliography{pore.bib}

\end{document}